\documentclass[aps,prl,superscriptaddress,twocolumn,showpacs]{revtex4}
\usepackage{amsmath}
\usepackage{graphicx}
\usepackage{amssymb}
\usepackage{bbm, color, soul}

\newcommand*{\tn}[1]{{\textnormal{#1}}}

\usepackage{color}

\begin{document}

\title{Behavior of Shannon entropy around an exceptional point in an open microcavity
}

\author{Kyu-Won Park}
\affiliation{School of Physics and Astronomy, Seoul National University, Seoul 08826, Korea}
\author{Jinuk Kim}
\affiliation{School of Physics and Astronomy, Seoul National University, Seoul 08826, Korea}

\author{Songky Moon}
\affiliation{School of Physics and Astronomy, Seoul National University, Seoul 08826, Korea}

\author{Kyungwon An}
\email{kwan@phya.snu.ac.kr}
\affiliation{School of Physics and Astronomy, Seoul National University, Seoul 08826, Korea}

\date{\today}

\begin{abstract}
 We have investigated the Shannon entropy around an exceptional point (EP) in an open elliptical microcavity as a non-Hermitian system. The Shannon entropy had an extreme value at the EP in the parameter space. The Shannon entropies showed discontinuity across a specific line in the parameter space, directly related to the occurrence of exchange of the Shannon entropy as well as the mode patterns with that line as a boundary. This feature results in a nontrivial topological structure of the Shannon entropy surfaces.
\end{abstract}
\maketitle

Understanding the characteristics of open physical systems has been a very essential and fundamental issue since there are always system-bath interactions in real physical systems. The convenient and effective ways to investigate this openness effect is to consider a non-Hermitian system~\cite{R09,M11}. The openness effects in a non-Hermitian system are significantly exhibited in the vicinity of a singular point called an exceptional point (EP), where not only the complex eigenvalues but also their eigenmodes coalesce~\cite{K66,W04}.

The investigations of physical effects near the EP have been extensively conducted in various areas such as atomic physics~\cite{HJ07,HM16}, microwave cavities~\cite{CB03,SB14}, photonic crystals~\cite{AA16,SM17}, optical microcavities~\cite{SJ09,WS17,SJ12}, ultrasonic acoustic cavities~\cite{SK16} and so on, both theoretically and experimentally. They have not only provided useful applications such as microcavity sensors~\cite{J14,HA17,S17} and enhancement of spontaneous emission~\cite{ZA16,Y06,AB17}, but also revealed many intriguing concepts and phenomena related to parity-time symmetry~\cite{LR17,SC15,RK18,aa14}, chirality~\cite{AG17,TG18,BS16,WG17}, phase transition~\cite{YW14,AH15,PJ00} and topological transfer of energy~\cite{HD16}. To the best of our knowledge, however, there have been no attempts to address the behavior of eigenmodes near the EP in the perspective of information theory. In this Letter, we challenge this task by introducing the Shannon entropy for the probability density of eigenmodes around an EP in a dielectric microcavity.

The Shannon entropy 
is defined as a measure of the average information content associated with a random outcome~\cite{C48}. Originally introduced in data communication~\cite{C48} and information theory~\cite{J91,RM92}, it is now utilized in diverse research fields. The Shannon entropy has been studied in association with the black holes~\cite{J73}, confined hydrogenic-like systems~\cite{WF18} and the entanglement~\cite{JS09} in physics. It was also applied to the bio-system~\cite{JW05,JC97}, the ecological modeling~\cite{SJ06} and information flow in finance~\cite{RH02}. Moreover, the Shannon entropy recently has also been used as an indicator for avoided crossing in dielectric microcavities~\cite{KS18} as well as atomic systems~\cite{GD033,HC15}.

To deal with the interactions in an open system, it is convenient to introduce a non-Hermitian Hamiltonian formulated by
\begin{align}
H=H_{\rm S}+V_{\rm SB}G_{\rm B}^{(\rm out)}V_{\rm BS},
\end{align}
where $H_{\rm S}$ is a Hermitian Hamiltonian for a closed system  (without interaction with a bath) associated with the open system, $G_{\rm B}^{(\rm out)}$ is an outgoing Green function in a bath, and $V_{\rm SB}$($V_{\rm BS}$) is the interaction from the bath (the closed system) to the closed system (the bath)~\cite{R09,M11}. It should be noted that the domain of $H$ is restricted to the part of the system excluding the bath, so are its eigenvectors~\cite{R09,K18}.
The matrix elements for $H$ are typically given by
\begin{align}
H=
\begin{pmatrix}
\epsilon_{1} & \omega \\
\omega & \epsilon_{2}
\end{pmatrix},
\end{align}
where $\epsilon_{i}\in \mathcal{C}\tn{(complex)}$, $\omega\in \mathcal{R}\tn{(real)}$, and its eigenvalues are
\begin{align}
E_{\pm}=\frac{\epsilon_{1}+\epsilon_{2}}{2}\pm Z
\end{align}
with $Z=\sqrt{\frac{(\epsilon_{1}-\epsilon_{2})}{4}+\omega^{2}}$. Here we assume $\omega$ to be a real value to simplify the relation between strong and weak interactions: there is a repulsion in the real part of the energy eigenvalue with a crossing in the imaginary part for $2\omega>|\mathrm{Im}(\epsilon_{1})-\mathrm{Im}(\epsilon_{2})|$ while there is a repulsion in the imaginary part with a crossing in the real part for $2\omega<|\mathrm{Im}(\epsilon_{1})-\mathrm{Im}(\epsilon_{2})|$. The former (latter) case corresponds to the strong (weak) interaction. Especially, the eigenvalue is degenerate when $Z=0$, corresponding to an EP. It is well-known that the EP is a singular point where the transition between the strong and the weak interactions takes place~\cite{W90,W00,SJ08}.

\begin{figure*}
\centering
\includegraphics[width=0.8\textwidth]{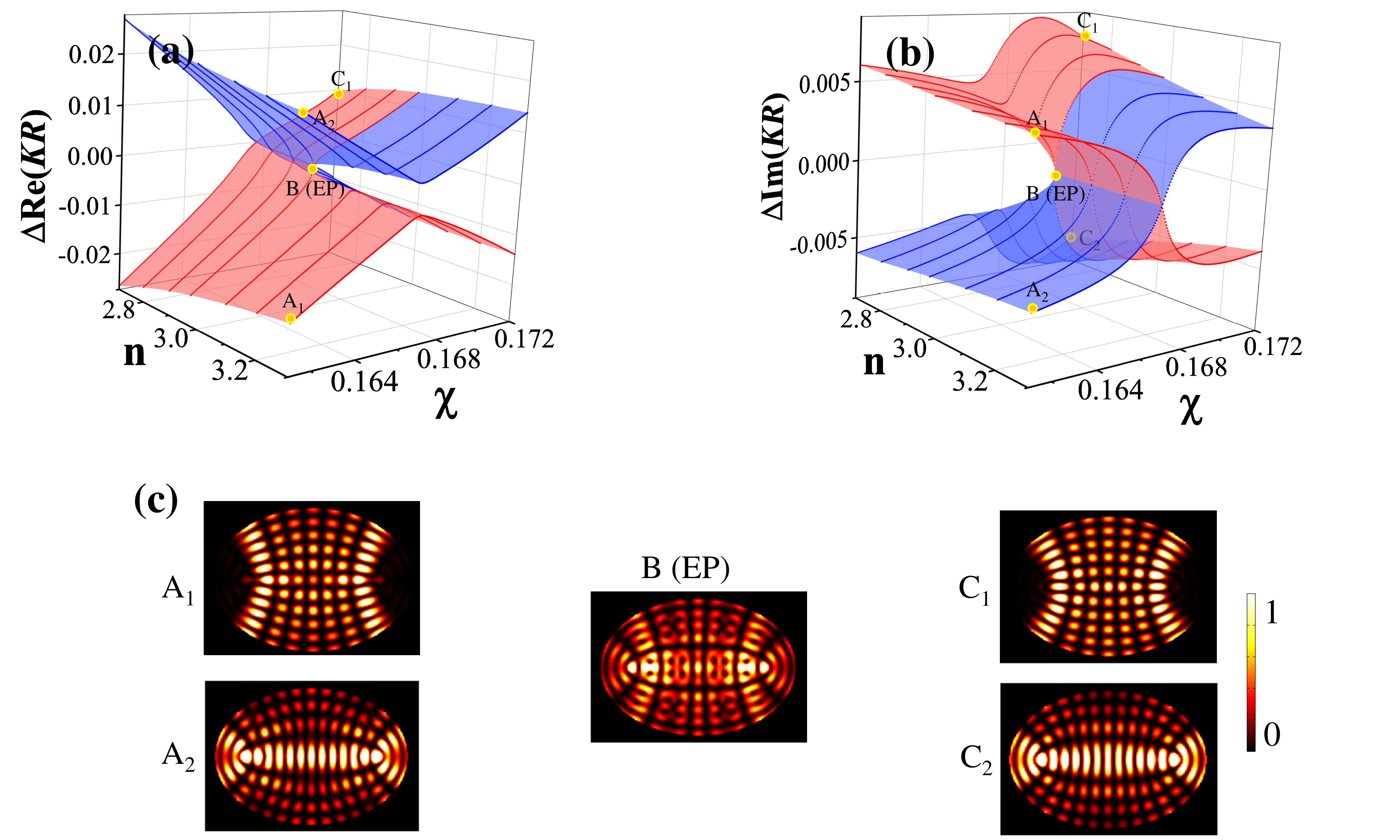}
\caption {(Color online) (a) The real parts of eigenvalues ($\Delta E\pm$) for a dielectric elliptical microcavity around an EP in the parameter space $(n,\chi)$. They show repulsions for $n\in(2.9777,3.3]$ (strong interaction regime) whereas showing crossings for $n\in[2.7,2.9777)$ (weak interaction regime).
(b) The imaginary parts of eigenvalues around the EP. On the contrary to the real parts, crossings occur for $n\in(2.9777,3.3]$ while repulsions for $n\in[2.7,2.9777)$. The EP is located at $(n_\tn{EP}\simeq 2.9777, \chi_\tn{EP}\simeq 0.16657)$.
(c) The mode patterns for two interacting modes $(A_{1,2}, B, C_{1,2})$ are plotted for $(n=3.3,\chi=0.161)$, $(n \simeq n_\tn{EP},\chi \simeq \chi_\tn{EP})$, and $(n=2.7,\chi=0.172)$, respectively. The mode patterns become the most uniform at the EP.
}
\label{Figure-1}
\end{figure*}

When an integrable billiard becomes open, the off-diagonal elements of the non-Hermitian Hamiltonian $H$ accounting for mode-mode interactions come only from the openness effect or from the external interaction ($V_{\rm SB}G_{\rm B}^{(\rm out)}V_{BS}$)~\cite{R09,K18}.
Therefore, we consider an elliptical dielectric microcavity as our open system in the present work.

Figure 1 depicts the eigenvalue surfaces ($\Delta E_{\pm}$) of the two interacting modes (shown in red and blue, respectively) around an EP in the parameter space $s=(n,\chi)$ for an elliptical dielectric microcavity.
We consider the eigenvalue differences, $\Delta E_{\pm}=E_{\pm}-E_{\rm AV}$ with $E_{\rm AV}=\frac{E_{+}+E_{-}}{2}$, from their average values $E_{\rm AV}$ instead of the eigenvalue themselves $E_{\pm}$ in order to display the EP structure clearly.
Here, $n$ is the refractive index of the cavity medium and $\chi$ is a deformation parameter associated with the major axis $a=R(1+\chi)$ and the minor axis $b=\frac{R}{1+\chi}$, respectively.  The eigenvalues are obtained with the boundary element method (BEM)~\cite{W03} 
for a transverse-magnetic (TM) mode and their values are presented in the size parameter $kR$ with $k$ the complex wave number.
In Fig.~\ref{Figure-1}, the two modes (red, blue) are divided by a reference line (n=2.9777), which separates the two regimes of interactions, {\it i.e.}, the strong and and weak interactions.
The EP is located at $\big(n_\tn{EP}\simeq2.9777, \chi_\tn{EP}\simeq 0.16657\big)$.

The mode patterns of two interacting modes at $(n=3.3,\chi=0.161)$ are labeled by ${\rm A}_{1,2}$, the mode pattern at $(n\simeq n_\tn{EP},\chi\simeq \chi_\tn{EP})$ by B (EP) and those at $(n=2.7,\chi=0.172)$ by ${\rm C}_{1,2}$. These mode patterns are plotted in Fig.~\ref{Figure-1}(c), respectively. Note that the mode pattern at B(EP) has more uniform probability than the others $({\rm A}_{1,2}, {\rm C}_{1,2})$.

\begin{figure}
\centering
\includegraphics[width=0.35\textwidth]{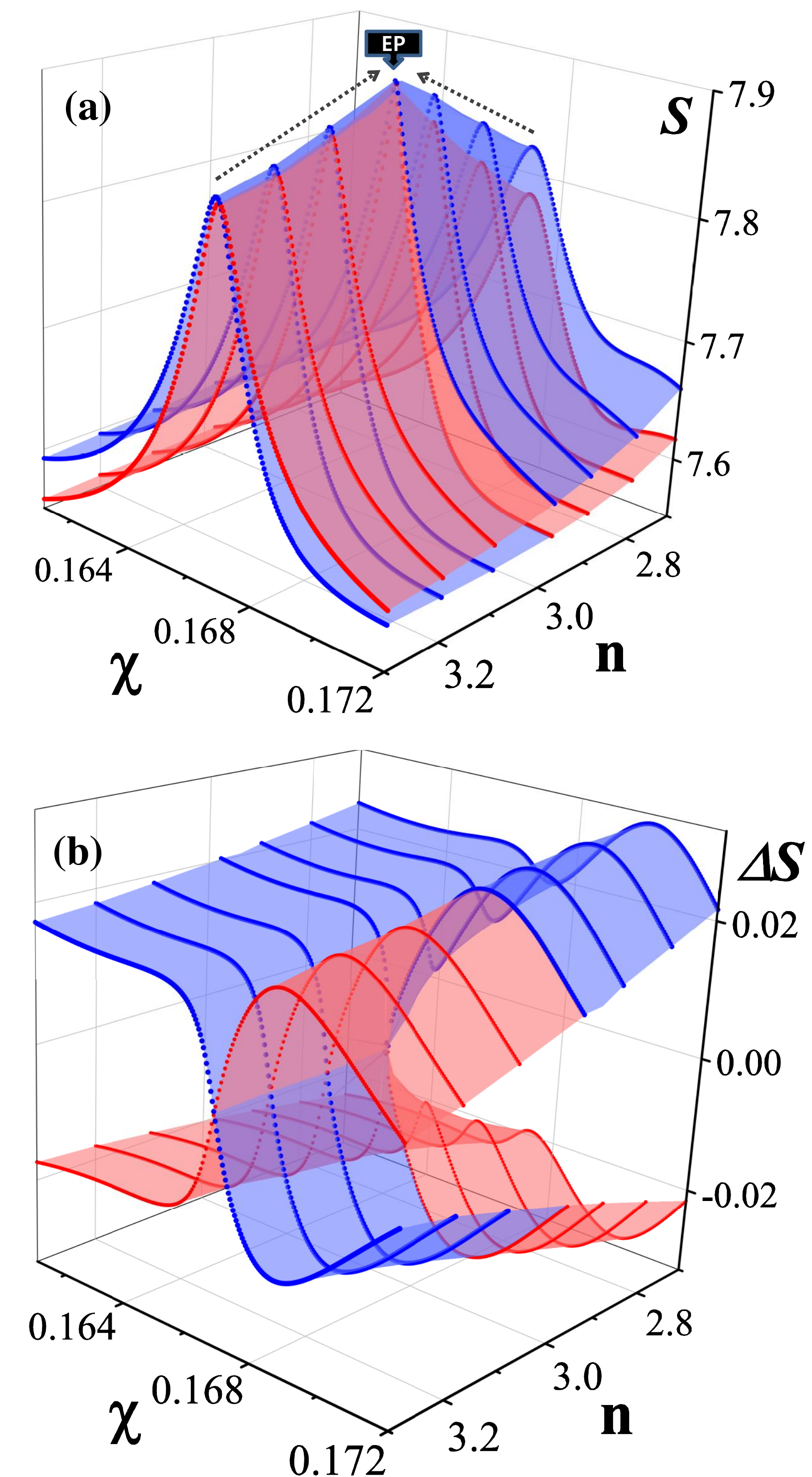}
\caption {(Color online) (a) The Shannon entropy for the intensity distributions of two interacting modes in a dielectric elliptical microcavity around an EP. The Shannon entropy is peaked at the center of interaction either in the strong or the weak interaction regime for a fixed refractive index $n$. These peaks have an extreme value $S(\rho)\simeq  7.8992$ at the EP. (b) The Shannon entropy $\Delta S(\rho)$ with respect to a mean is obtained in the same way as in Fig.~\ref{Figure-1}. The strurecture of $\Delta S(\rho)$ resembles that of $\Delta \tn{Im}(kR)$ in Fig.~\ref{Figure-1}(b).
}
\label{Figure-2}
\end{figure}

We now suggest that the Shannon entropy can be defined near an EP and can reveal the peculiar topology associated with the EP.
The Shannon entropy for a specific discrete probability distribution $\rho_i$ at $N$ number of different states is defined as
\begin{align}
S( \rho_i )= -\sum_{i=1}^{N}\rho_i \log\rho_i,
\end{align}
with a normalized condition $\sum_{i=1}^{N}\rho_i=1$. Here, we choose the mode intensity pattern inside the cavity as the probability distribution 
and the $N$-mesh points for the mode intensity pattern as the $N$ spatial-coordinate states of a fictitious particle in the corresponding billiard as our $N$ different states.
In Fig.~\ref{Figure-2}, the Shannon entropies of probability density for the two interacting modes around the EP in our elliptical microcavity $S(\rho; n,\chi)$ are plotted in the parameter space. The plots in Fig.~\ref{Figure-2} reveal two important features of Shannon entropy for EP, {\it i.e.}, an extreme value at EP and a nontrivial topological structure around it.

For the extreme value, we note that the Shannon entropy is maximized at the center of interaction at the fixed refractive index $n$ in both weak and strong interaction regimes. It is because the coherent superposition of eigenfunctions in either weak or strong interaction regime leads to an increase in Shannon entropy. More interestingly, the dotted black arrows in Fig.~\ref{Figure-2}(a) indicate that the trace of these maximum points has the extreme value at the EP with a value of $S(\rho)\simeq  7.8992$.

For the nontrivial topological structure, we observe that the two cyclic variations are required for the Shannon entropy values to return to the original values on the Shannon entropy surface, just like the complex eigenvalues on the complex energy surfaces. To see a clear connection between these two, we define $\Delta S_{1,2}\equiv S_{1,2}-S_{\rm AV}$ with $S_{\rm AV}=\frac{S_{1}+S_{2}}{2}$, similarly to $\Delta E_{\pm}$ in Fig.~\ref{Figure-1}. It is easily seen that the Shannon entropy surface resembles the imaginary part of the complex energy surfaces.

\begin{figure}
\centering
\includegraphics[width=0.48\textwidth]{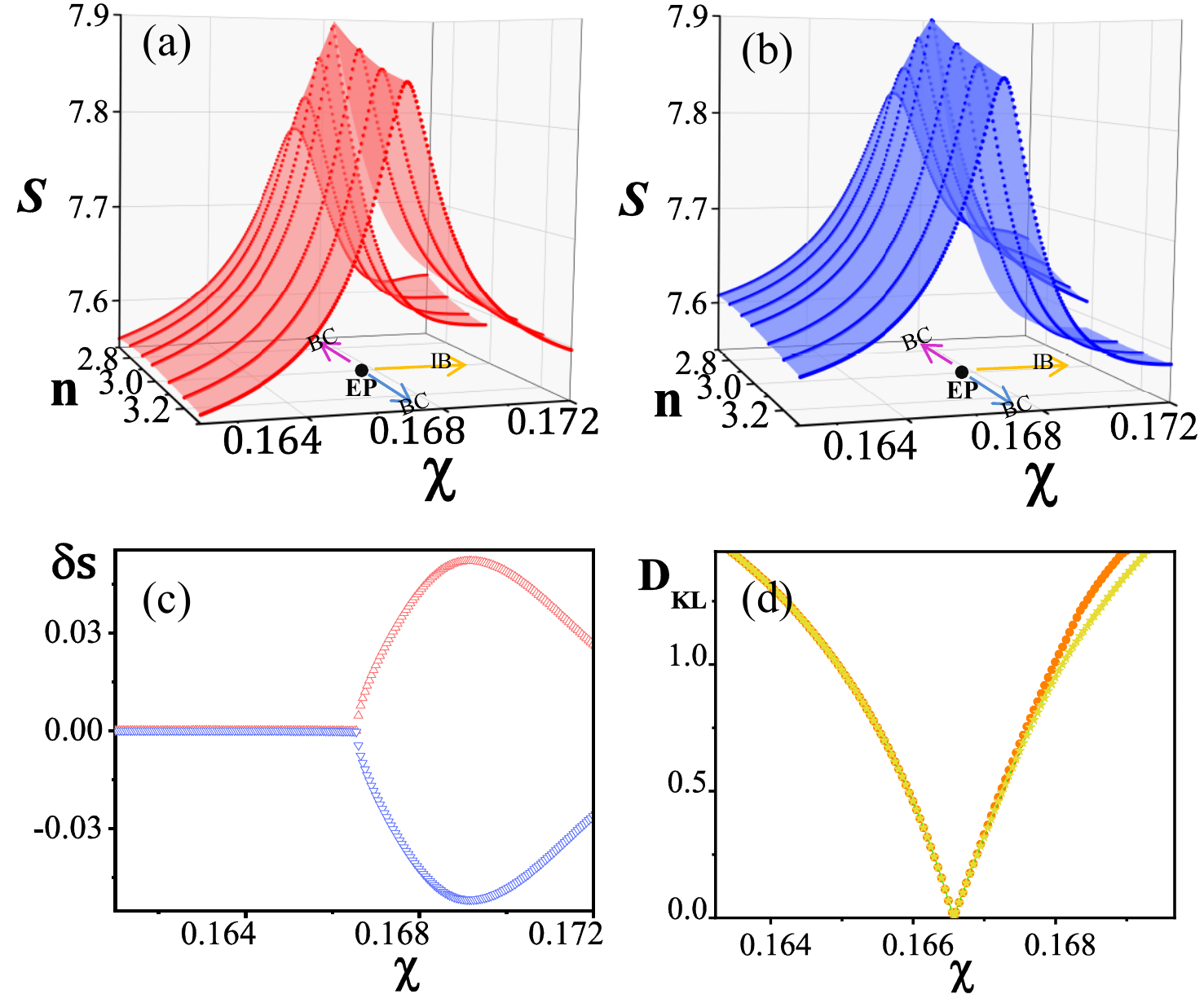}
\caption{(Color online) (a) and (b) show the Shannon entropies in Fig.~\ref{Figure-2} individually. The discontinuity appears along the line $n\simeq  n_\tn{EP}$ in both cases. (c) The difference $\delta S(\rho)$ of the Shannon entropies at $n_{-}$ and $n_{+}$. (d) The KL divergence $D_{\rm KL}$ or the relative entropy along $n_{\pm}$ for two interacting modes.
}
\label{Figure-4}
\end{figure}


 In order to investigate the origin of the nontrivial topological structure of the Shannon entropy surfaces, let us consider the Shannon entropy for each mode as shown in Fig.~3(a) and (b), respectively. The surface discontinuity is exhibited along the line $n\simeq  n_{\rm EP}$ in both cases. The discontinuity can be quantified by $\delta S(\rho)=S(\rho;n_{-},\chi)-S(\rho;n_{+},\chi)$, where $n_{\pm}=n_\tn{EP}\pm \delta n$, with $\delta n=0.0001$, for example. The result is shown in Fig.~3(c), where
$\delta S(\rho)$ remains almost zero for $\chi<\chi_\tn{EP}$ whereas it increases significantly for $\chi>\chi_\tn{EP}$ along the line $n \simeq n_{\rm EP}$.
This line is where the two interacting modes (red, blue) on different branches merge together, so let us call this line the interaction branch. A schematic diagram for the EP (branch point), the branch cut (BC) and the interaction branch(IB) is shown on the base planes in Fig.~3(a) and (b), respectively.

The discontinuity across the interaction branch is directly related to the exchange of the Shannon entropy as well as the mode exchange. This observation is consistent with our previous work~\cite{KS18}, where the Shannon entropy is exchanged with the repulsion in the real part of eigenvalues in the strong interaction regime.
This exchange property can be quantified by introducing the relative entropy. The relative entropy or the Kullback-Leibler (KL) divergence between the two probability distributions on a random variable is a measure of the distance between them~\cite{KL51}.
The KL divergence from $Q$ to $P$, usually denoted by $D_\tn{KL}\big(P\parallel Q\big)$, is defined by
\begin{align}
D_\tn{KL}(P\parallel Q)= -\sum_{i=1}^{N}P(r_{i})\log \frac{Q(r_{i})}{P(r_{i})}.
\end{align}
The KL divergence for the two interacting modes along the $n_\pm$ lines, respectively, is plotted in Fig.~3(d). The yellow (orange) symbols represent the KL divergence $(D^{w,(s)}_\tn{KL}(P\parallel Q))$ in the weak (strong) interaction regime at $n_{-}(n{+})$ as the $\chi$ is varied. It is seen that the KL divergences are almost degenerate when $\chi<\chi_{\rm EP}$ and they become zero at the EP: $\Delta D^{w,s}_\tn{KL}\equiv \left| D_\tn{KL}(P^{w}\parallel Q^{w})-D_\tn{KL}(P^{s}\parallel Q^{s})\right|\simeq  0$ when $\chi<\chi_\tn{EP}$ and $D^{w,s}_\tn{KL}=0$ at the EP.
However, the difference $\Delta D^{w,s}_\tn{KL}$ becomes larger across the interaction branch when $\chi>\chi_\tn{EP}$.
These results are consistent with the fact that the mode patterns as well as the Shannon entropies in the weak interaction regime are not exchange
whereas those in the strong interaction regime are exchanged.
The transition from mode-pattern non-exchange to mode-pattern exchange 
gives rise to the intersection of the Shannon entropy surfaces as seen in Fig.~\ref{Figure-2} and leads to the nontrivial topological structure of the Shannon entropy in the parameter space.

In summary, we proposed the Shannon entropy for investigating the behavior of eigenmode patterns near an EP in a dielectric elliptical microcavity in the perspective of the information theory. Our study yielded two interesting results.
First, the Shannon entropy has the extremal value at the EP.  Second, the Shannon entropy surfaces show a nontrivial topological structure with two cyclic variations and this feature analyzed with the relative entropy is associated with a discontinuity across the interaction branch in the parameter space. 

We thanks Sera Yu for useful comments. This work was supported by Samsung Science and Technology Foundation under Project No. SSTF-BA1502-05, the Korea Research Foundation (Grant No.~2016R1D1A109918326) and the Ministry of Science and ICT of Korea under ITRC program (Grand No.~IITP-2019-0-01402).

\end{document}